\documentstyle [12pt, epsfig]{article}
\begin{document}
 
\hfill {WM-02-101}

\hfill {\today}

   \baselineskip 24pt

{
\Large
   \bigskip
   \centerline{Large Electric Dipole Moments of Heavy Neutrinos }
 }

\vskip .2in
\def\bar{\overline}

\centerline{Marc Sher \footnote{Email: sher@physics.wm.edu}  }
\centerline {\it Nuclear and Particle Theory Group}
\centerline {\it Physics Department}
\centerline {\it College of William and Mary, Williamsburg, VA 23187, USA}
\bigskip
\centerline{Shuquan Nie \footnote{Email: sxnie@alcor.concordia.ca} }
\bigskip
\centerline {\it Physics Department, Concordia University, 1455 De Maisonneuve Blvd. W.}
\centerline {\it Montreal, Quebec, Canada, H3G 1M8}

\vskip 0.5in

{\narrower\narrower  In many models of CP violation, the electric dipole moment (EDM) of a heavy charged or
neutral lepton could be
very large.  We  present an explicit model in which a heavy neutrino EDM can be as large as $10^{-16}$ e-cm,
or even a factor of ten larger if fine-tuning is allowed, and use an effective field theory argument to show that
this result is
fairly robust.   We then look at the production cross section for these neutrinos, and by rederiving the
Bethe-Block formula, show
that they could leave an ionization track.  It is then noted that the first signature of heavy neutrinos with a
large EDM would
come from 
$e^+e^-\rightarrow
\bar{N}N\gamma$, leading to a very large rate for single photon plus missing energy events, and the rate and
angular distribution
are found.   Finally, we look at some astrophysical consequences, including whether these neutrinos could
constitute the UHE cosmic
rays and whether their decays in the early universe could generate a net lepton asymmetry.}

\newpage
\section{Introduction}

Most of the unknown parameters of the Standard Model come from fermion masses and mixing angles.  They are put
in by hand, and we have no real understanding of their origins.   Many models that try to explain their values
exist, most involving additional symmetries, but more experimental data will be needed to distinguish between
them.   Much of the excitement concerning neutrino masses and mixing angles is caused by the hope that their
values will facilitate our understanding of the flavor problem.
 
A potentially valuable source of information about the flavor problem may come from electric and magnetic
dipole moments.  Just as the Yukawa couplings form a matrix in generation space, the interaction of two
fermions with a photon will also be a matrix in generation space.  The real and imaginary parts of the diagonal
elements will lead to the magnetic and electric dipole moments; the off-diagonal elements wil lead to
flavor-changing radiative decays.   If one can measure these moments, one would obtain valuable hints of
physics beyond the Standard Model and would learn more about the origin of flavor.

This paper is concerned with the electric dipole moments (EDMs) of leptons.  There is a possibility of
substantial improvement in the experimental bounds on EDMs in the near future.  Proposals exist\cite{muonedm} to
lower the current bound\cite{muoncurrent} on the muon's EDM by six orders of magnitude, and substantial
improvement in the bound on the electron EDM\cite{electroncurrent} is possible.   Combining limits on the weak
dipole moment of the tau with $U(1)$ symmetry\cite{masso} improves the current bound\cite{taucurrent} by a
factor of thirty.

How big might one expect the EDMs to be?  In the Standard Model, they are negligibly small\cite{currentedms}. 
However, they can be much larger in extensions of the Standard Model.  In multi-Higgs models, the EDM of the
muon can easily be as large as $10^{-24}$ e-cm, within reach of planned experiments\cite{barger}.  In leptoquark
models, the muon and tau EDM's are typically $10^{-24}$ e-cm and $10^{-19}$ e-cm, respectively\cite{bbo}. 
Left-right models\cite{left-right} have a muon EDM which is typically of the order of $10^{-24}$ e-cm, and in
the minimal supersymmetric standard model(MSSM)\cite{mssm}, the electron EDM is above the experimental bounds if the
phases are all unity.  The point is that a wide variety of models give EDMs that can be observed in the near
future.

Babu, Barr and Dorsner\cite{bbd} have discussed how the EDMs of leptons scale with the lepton masses.  In many
models, such as the Standard Model and the MSSM, they scale linearly with the mass.  However, in a large number
of models, such as multi-Higgs, leptoquark and some flavor symmetry models, the EDM scales as the cube of the
lepton mass.  In these models, the tau EDM will be 5000 times larger than that of the muon (it should be noted
that the electron EDM in some of these models receives a two-loop contribution which varies only linearly, and
thus it need not be negligible).

Given cubic scaling, the EDM of a heavy lepton (charged or neutral) could be quite large.  For example, a $100$
GeV heavy lepton would have an EDM a billion times larger than the muon's.  If the latter is in the expected
range of $10^{-24}$ e-cm, then such a heavy lepton would have an EDM as large as $0.01$ e-fermi.   This EDM
would then dominate the electromagnetic interactions of these leptons, drastically changing their
phenomenology.  In particular, if a heavy neutrino acquired such an EDM, it could even leave an observable
ionization track.

Although originally motivated by cubic scaling models, the possibility of a heavy lepton having a huge EDM is
worth studying in a model-independent way.  We will see below that plausible models exist in which a heavy
neutrino has a very large EDM, even though these models do not have cubic scaling.   In this Article, we will
concentrate on the possibility that a heavy neutrino could have a large EDM.   In an earlier paper by one of
us\cite{sher}, the differential production cross section for heavy leptons (charged and neutral)  with large EDMs
was studied, and
the fact that heavy neutrinos with large EDMs could leave an ionization track was discussed (although no explicit
models were
mentioned).    This Article extends this earlier work substantially.  In  Section II, we explicitly present a
model in which the
EDM of a heavy neutrino can be of $O(10^{-16})$ e-cm.   This will serve as an existence proof that relatively
simple models can
exist in which this occurs.  Current bounds on such large EDMs
will be noted.   In Section III, the cross section for heavy neutrino production is presented, and the
Bethe-Block formula for a large-EDM heavy neutrino traversing matter will be calculated.  This extends slightly
previous work.  The result will be that an ionization track is observable, but one would need to either modify
existing detectors or construct a new detector.   Since it is unlikely that this would occur without some
preliminary evidence that such a large-EDM neutrino exists, we look, in Section IV, at the process of initial and final
state radiation,
$e^+e^-\rightarrow N\bar{N}\gamma$, which would give an enormous rate for single-photon plus missing energy
events, and would provide impetus for looking explicitly for large-EDM heavy neutrinos.  Finally, in Section
V, we discuss whether these heavy neutrinos could be candidates for the ultra-high energy cosmic rays, and also
consider the implications for baryogenesis through leptogenesis, and in Section VI, we present our conclusions.

\section{Plausible Models}

Are there plausible models in which a huge EDM for a heavy neutrino occurs?  If one does not rely on any
particular model, then a simple effective field theory argument can be used.   Suppose one assumes that $CP$
violation is due to some sort of new physics at the TeV scale.   Then one can write the effective low-energy,
dimension-five Lagrangian as
\begin{equation}
{\cal L}={c\over \Lambda}\bar{L}_L\sigma_{\mu\nu}i\gamma_5 L_R F_{\mu\nu}.
\end{equation}
If $\Lambda$ is $O(1)$ TeV, and the unknown coefficient is of $O(1)$, then this yields a very large EDM of
approximately $10^{-15}$ e-cm for the lepton.

Of course, in realistic models, one expects the EDM to occur via a loop, suppressing the unknown coefficient,
and there may not even be a dimension-five operator.   What about specific models currently in the literature? 
A model with cubic scaling is that of Bernreuther, Schroder and Pham\cite{bernreuther}, in which CP violation
occurs in the Higgs sector.   In this model, the EDM of a heavy lepton (they consider the top quark, but the
results are unchanged) is constrained by the electron EDM, and the maximum value is
$O(10^{-17})$ e-cm.   In other models, such as the model of Babu, Barr and Dorsner\cite{bbd} in which their
parameter $c=0$, the electron EDM does not give any such constraints and a large EDM is allowed.

Here, we will provide an explicit example of a model with a large EDM.  Consider a model containing singlet
fermions, $E_{L,R}$ and $N_{L,R}$ and two charged singlet scalar fields, $h_i$.  The fermions can  couple to the
charged
scalars,
$h_i,
\ i$=1, 2, in the following form
\begin{equation}
Y_i \bar{E}_R N_L h_i +Z_i \bar{E}_L N_R h_i + h.c.
\end{equation}
where $Y_i$ and $Z_i$ are complex Yukawa couplings and can generally  
include flavor indices. The $h_i$ can also mix with each other and can be diagonalized
by a complex unitary matrix U: $h_i=U_{ij} H_j$, with mass eigenstates $H_i$.
Two such charged singlets are necessary so that a phase redefinition cannot eliminate the CP violation.
For simplicity in presentation, we will look at the single $H$ case, and just assume that the phase redefinition
does not occur.   In this case, the coupling is
\begin{equation}
\bar{E}(S-P \gamma_5 )N H + h.c.
\end{equation}     
where S and P are arbitrary complex couplings.
 
\begin{figure}
\centerline{ \epsfysize 1.5in \rotatebox{360}{\epsfbox{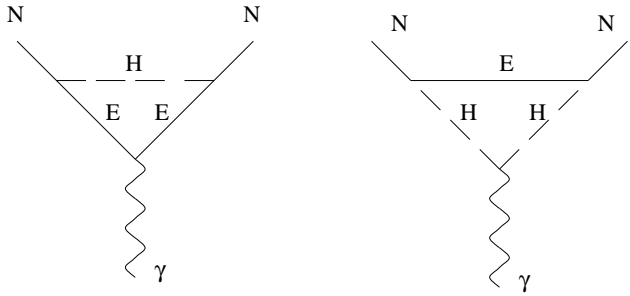}}  }
\caption{One-loop feynman diagrams contributing to a large EDM for a heavy neutrino. }
\protect \label{figEDM}
\end{figure}

It is straightforward to calculate the EDM due to the above interaction. The one-loop
feynman diagrams are listed in Fig. \ref{figEDM}.  The EDM is defined at $q^2=0$, and is given by
\begin{equation}   
d_N(q^2=0) = e \frac{m_E}{8 \pi^2} Im(SP^*) \int_0^1 dz \frac{(1-z)(1+2z)}{-z(1-z)m_N^2+(1-z)m_E^2+z m_H^2}
\end{equation}
which, numerically, in units of e-cm, is
\begin{equation}
3.2\times 10^{-17}\left({100 {\rm GeV}\over M_E}\right) {Im(SP^*)\over 4\pi}
\int_0^1 dz \frac{(1-z)(1+2z)}{-z(1-z)a^2+(1-z)+z b^2}
\end{equation}
where $a\equiv M_N/M_E$ and $b\equiv M_H/M_E$.  

\begin{figure}
\centerline{ \epsfysize 3.0in \rotatebox{360}{\epsfbox{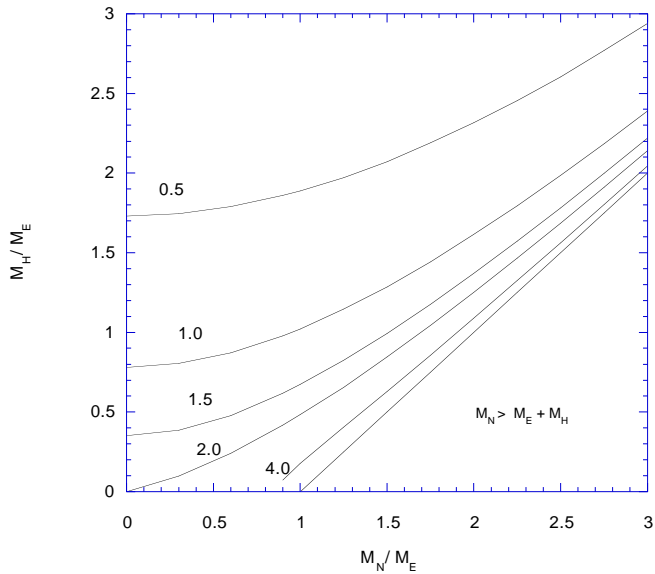}}  }
\caption{Value of the integral in the expression for the EDM. }
\protect \label{figintegral}
\end{figure}

The integral is plotted in Fig. \ref{figintegral}.  We see that EDM's of $10^{-16}$ e-cm are quite possible,
especially in the region of paramter-space in which the $E$ is heavier than the $N$ or $H$, if one chooses
${Im(SP^*)\over 4\pi}\sim 1$,  and can even be an order of magnitude higher if one allows for some fine-tuning.
We have not considered the unlikely region of parameter space in which $M_N > M_E + M_H$, since the width of the
$N$ in this case is comparable to its mass, and the definition of the EDM is no longer straightforward.

We thus have presented a  model in which the EDM of a heavy neutrino is very large, of $O(10^{-16})$ e-cm. 
While this model is, admittedly, designed to have such a large EDM, its ingredients are very simple, and one can
imagine that it may be part of a more complicated model.  The main point is that having a heavy neutrino with a
large EDM is certainly not excluded, and in this paper we are exploring the phenomenological implications.

\begin{figure}
\centerline{ \epsfysize 1.5in \rotatebox{360}{\epsfbox{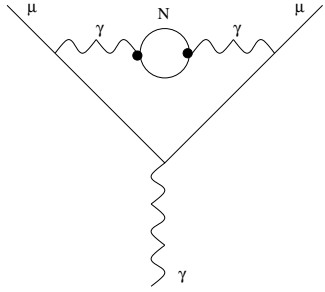}}  }
\caption{Large EDM contribute to $a_{\mu}$}
\protect \label{figg-2}
\end{figure}

Are there any current phenomenological bounds?  If the heavy neutrino and charged lepton are part of a
fourth chiral fermion family (albeit with a  right-handed neutrino), then the S and T parameters will cause
severe constraints, but if they are part of a vectorlike multiplet, there will be no such constraints.  Very
explicit models may be constrained, but there are no general bounds.  One possible concern would be g-2
of the muon.   Here, there will be an effect of a large EDM on the photon propagator. At the lowest order, the
feynman  diagram is illustrated in Fig. \ref{figg-2}. As in the calculation of
hadronic contributions to $(g-2)_{\mu}$, we calcualte the effect of 
a large EDM on $(g-2)_{\mu}$ by the dispersion integral \cite{Czarnecki}  

\begin{equation}
a_{\mu}^{EDM}=\frac{1}{4 \pi^3} \int_{4 m_N^2} ds K(s) \sigma(s)_{e^+e^- \rightarrow NN},
\end{equation}
where K(s) is given in Ref. \cite{Czarnecki} and $\sigma(s)_{e^+e^- \rightarrow NN}$ is given in 
Ref. \cite{sher}, as well as below.  We cutoff the upper bound of the integral. Letting $m_N=100$ GeV, 
$a_{\mu}^{EDM}=5.9 \times 10^{-11} \ (1.3 \times 10^{-10})$ for a cutoff of $1 \ (10)$ TeV. 
This is much smaller than the theoretical and experimental uncertainties, and is thus negligible.
  
\section{Direct detection}

The most dramatic effect of a large EDM of a heavy neutrino will be in the 
production cross-section and angular distribution.   Escribano
and Masso\cite{masso} noted that the relevant $U(1)$ invariant effective operator  is
given by $\bar{L}_L\sigma^{\mu\nu}i\gamma_{ 5}L_RB_{\mu\nu}$, where $B_{\mu\nu}$ 
is the $U(1)$ field tensor.  This gives a coupling to the photon, 
which we define to be the EDM, as well as a coupling to the $Z$ which is the EDM times $\tan\theta_W$.  We will
include this coupling to the $Z$.  It turns out that the contribution from the $Z$ has very little effect on the
numerical results.   In general, one should also include
an operator coupling to the $SU(2)$ field tensor, leading to a very different
value for the CP-violating coupling to the $Z$.  Rather than deal with two parameters, however, we just
assume that the latter operator is smaller.  If this assumption is false, then it will  just make the
cross-section even bigger, unless there is fine-tuning.  Also, unless the CP-violating coupling to the $Z$ is
surprisingly large, it will have very  little effect on the results.

A discussion of the differential cross section for a heavy charged lepton can be found in Ref. \cite{sher}. 
Here we are interested in heavy neutrino production.  
The differential cross-section is given by
\begin{equation}
{d\sigma\over d\Omega}={\alpha^2\over 4s}\sqrt{1-{4M^2\over s}}\left( A_1 + 
{1\over 8\sin^42\theta_W} P_{ZZ}\ A_2 +
{(1-4\sin^2\theta_W)\tan\theta_W\over \sin^22\theta_W} P_{\gamma Z}\ A_3\right)
\end{equation}
where 
\begin{eqnarray} A_1&=&
D^2s\sin^2\theta(1+{4M^2\over\
 s})\nonumber \\ A_2&=&
1+\cos^2\theta -{4M^2\over s}\sin^2\theta + 8C_V\cos\theta+
D^2s\tan^2\theta_W(\sin^2\theta +{4M^2\over s}(1+\cos^2\theta))\nonumber \\
A_3&=&4D^2s(\sin^2\theta+
{4M^2\over s}(1+\cos^2\theta))
\end{eqnarray}
where we have dropped the numerically negligible $C_V^2$ terms, for simplicity.

\begin{figure}
\begin{center}
\ \hskip -2cm \epsfysize 3in\epsfbox{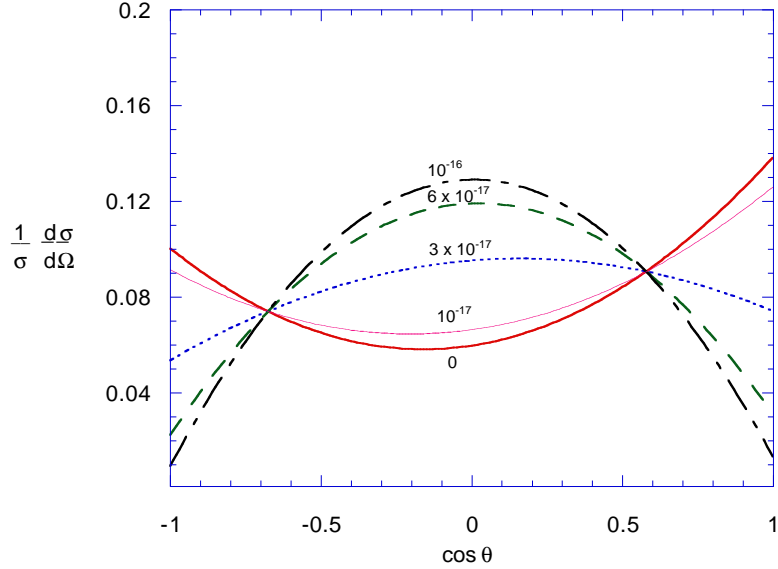}
\caption{Differential cross section for heavy neutrino production for various EDMs, in
units of e-cm, for a lepton mass of 100 GeV.}
\end{center}
\end{figure}

\begin{figure}
\begin{center}
\ \hskip -2cm \epsfysize 3in\epsfbox{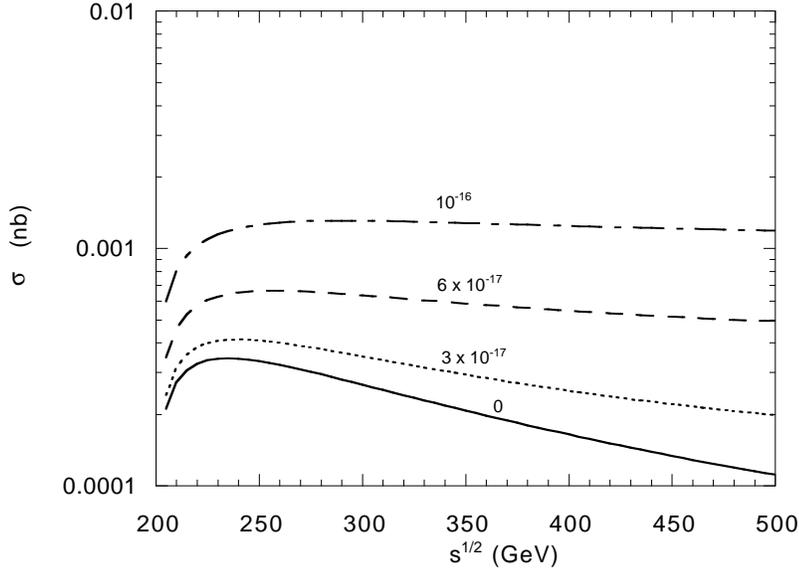}
\caption{The total cross section for heavy neutrino production for various EDMs, in
units of e-cm, for a lepton mass of 100 GeV.}
\end{center}
\end{figure}
The differential and total cross-sections are given in Figures 4 and 5, for 
a heavy neutrino mass of $100$ GeV.   We see that for $D=0$, the usual
$1+cos^2\theta + C \cos\theta$ distribution for a lepton (where the $\cos\theta$ term is due to 
$\gamma-Z$ interference) is found.  For a very large EDM,  the distribution is completely dominated by the
$\sin^2\theta$ contribution from the electric dipole term.   The angular distribution begins to deviate
significantly from the $D=0$ distribution for $D$ greater than
$3\times 10^{-17}$, and by $D=3\times 10^{-16}$ the distribution is very close to $\sin^2\theta$.   
One can see that EDM's of the size noted earlier will dramatically alter the angular distribution\cite{sher}.

Note that since the cross-section varies as $D^2$,  for an EDM as ridiculously large as
$1.0$ e-fermi, it would be almost a microbarn!!  Cross sections this
large will violate unitarity.  The unitarity limit can be 
estimated by setting the relevant effective interaction strength, $\alpha D\sqrt{s}$, equal to unity.  For
 $\sqrt{s}=500$ GeV, $D=10^{-15}$ e-cm, 
and $\alpha\simeq 1/125$, this effective interaction strength is $0.25$.  Thus
the unitarity limit will not be reached unless the EDM is larger than $10^{-15}$ e-cm, and so the behavior discussed above will
appear for a wide range of parameter space without violating the unitarity bound.  Another way of saying this is to note that 
the larger the EDM, the smaller the scale at which the physics responsible for
the effective interaction sets in, and for an EDM larger than 
$10^{-15}$ e-cm, that scale is less than $\sqrt{s}$.

Production of a fermion with a large EDM has been considered elsewhere.  It is given in the context of a top
quark EDM by Bernreuther et al.\cite{bernreuther}, and given in the context of tau-pair 
production in Ref. \cite{bno}.  This latter paper noted how one can use CP-odd 
angular correlations to search for a tau EDM, and this method has been used
by experimentalists.  However, there has not been any discussion of the 
possibility of a large EDM for heavy neutrinos, and here we see a unique and unusual
signature.

What is the purpose of this calculation?  After all, it would not seem remotely possible to detect heavy
neutrinos directly.  This calculation would only be meaningful if the heavy neutrino were heavier than the
charged lepton, and could thus be detected via its decay.  Here, one could again look for CP-odd correlations,
as discussed in Ref. \cite{bno}.   However, most models have the charged
lepton heavier than the neutrino, and thus the decay can only occur through
mixing with the very light neutrinos.   As discussed in detail in Ref. \cite{Frampton}, this mixing could be
very  small, and these heavy neutrinos could be
effectively stable. As we will see below, however, 
it may be possible to detect these neutrinos initially through final state radiation, and then directly.

Because the heavy neutrino has such a large electric dipole moment, it might be possible to detect it directly
through its ionization loss.  To estimate  whether or not this might be possible, we follow the derivation of
the classical Bohr formula outlined in Jackson\cite{jackson}, replacing the electric field from a charge with
the electric field from a dipole.  This is a classical calculation; if one is not too far above threshold, the
neutrino will be moving non-relativistically.  Consider a heavy neutrino moving in the $x$-direction, and an
electron  at an impact parameter $y=b$.  The impulse given to the electron is
$\Delta\vec{p}=\int_{-\infty}^\infty \ e\vec{E}dt$.   The result will depend on the orientation of the dipole. 
Suppose that the dipole is in the $z$-direction, transverse to the plane of the particle motion and the
electron.  Then the only nonzero electric field component is in the $z$-direction, and $E_z={eD\over
4\pi\epsilon_o} (b^2+v^2t^2)^{-3/2}$, where the time of closest approach is defined to be $t=0$, $v$ is the
velocity, and $eD$ is the size of the electric dipole moment.  Integrating, one finds that the impulse is
${e^2D\over 4\pi\epsilon_o}{2\over vb^2}$.   Suppose that the dipole is in the $y$-direction.  Then the
electric field components are $E_x={eD\over 4\pi\epsilon_o r^3}(3\sin\theta\cos\theta)$ and $E_y={eD\over
4\pi\epsilon_o r^3}(3\cos^2\theta-2)$, where $r^2=b^2+v^2t^2$ and $\tan\theta=b/vt$.    Integrating, the
impulse in the $x$-direction vanishes, as expected by symmetry, and the impulse in the $y$-direction is
also ${e^2D\over 4\pi\epsilon_o}{2\over vb^2}$.  Finally, if the dipole is in the $x$-direction, the electric
field components are $E_y={eD\over 4\pi\epsilon_o r^3}(3\sin\theta\cos\theta)$ and $E_x={eD\over
4\pi\epsilon_o r^3}(3\cos^2\theta-1)$.   Both of these integrate to zero, so there is no net momentum transfer
in this case.   Since the orientation is generally arbitrary, and we are only interested in a rough
order-of-magnitude estimate, we will take the impulse to be $|{\Delta \vec{p}}|={e^2D\over
4\pi\epsilon_o}{2\over vb^2}$. Note that the impulse from an electric charge is just ${2e^2\over 4\pi\epsilon_o
vb}$, so this
result is expected dimensionally.   This impulse is then converted into an energy transfer, $\Delta E=|\Delta
\vec{p}|^2/2m$.  Jackson notes that the maximum energy transfer is $\Delta E_{max}=2m\gamma^2v^2$, and thus
the minimum impact parameter is $b^2_{min}=e^2D/(m\gamma v^2)$.   The energy loss is obtained by cylindrically
integrating over the impact parameter
\begin{equation}
{dE\over dx}=2\pi NZ\int_{b_{min}}^\infty \Delta E(b) b db
\end{equation} 
and we obtain
\begin{equation}
{dE\over dx}= 2\pi N_A\left({e^2\over 4\pi\epsilon_o}\right) D\gamma {Z\over A}
\end{equation}
which is the corresponding formula to the Bohr result.  Since we are
doing a classical calculation, one can set $\gamma=1$.  Note that the logarithm in the usual Bethe-Block
formula is absent and the electron mass and particle velocity drop out.  This is due to the extra power of $b$
in the expression for the impulse, which arises on dimensional grounds.  Plugging in numbers, one finds that
${dE\over dx}=10^{12}D \ {\rm MeV} {\rm g}^{-1} {\rm cm}^2$.

For EDM's of $10^{-16}$ e-cm, this gives an energy loss which is approximately $10^{-4}$ of the usual energy
loss for a charged particle.  This is challenging for experimenters, but not impossible, requiring specialized
detectors. For example, if the heavy neutrino is produced in the decay of a charged lepton, then it would be
produced in coincidence with a real or virtual W, which would eliminate backgrounds.   Note that the event
rate, for $D=10^{-16}$ cm, is huge, corresponding to $100,000$ events for an integrated luminosity of $100\
{\rm fb}^{-1}$.  This is several events per hour, and thus a specialized detector located on the outside of the
main detectors could suffice, without needing a full $4\pi$ coverage (note that the $N$'s will emerge
back-to-back).  More importantly, however, is that initial and final state radiation  will be
huge, and thus standard searches for single-photon plus missing energy events will discover a huge signal,
which could then be followed by a specialized search for direct detection.  We now turn to the
possibility of detection through single photon events.

\section{Detection through single photon events}

In the standard model, the process $e^+e^- \rightarrow \nu \bar{\nu} \gamma$
has been calculated and used to determine the number of light neutrinos 
\cite{Ma, Gaemers}. The signature of the process is a single
photon from initial state radiation plus missing energy. The process didn't rule out the existence of heavy
neutrinos. 
Similarly, the process $e^+e^- \rightarrow N \bar{N} \gamma$ can be used to study the effects
of large EDM of heavy neutrinos. Now the single photon can also come from final state radiation
because of the large EDM.    

\begin{figure}
\centerline{ \epsfysize 3in \rotatebox{360}{\epsfbox{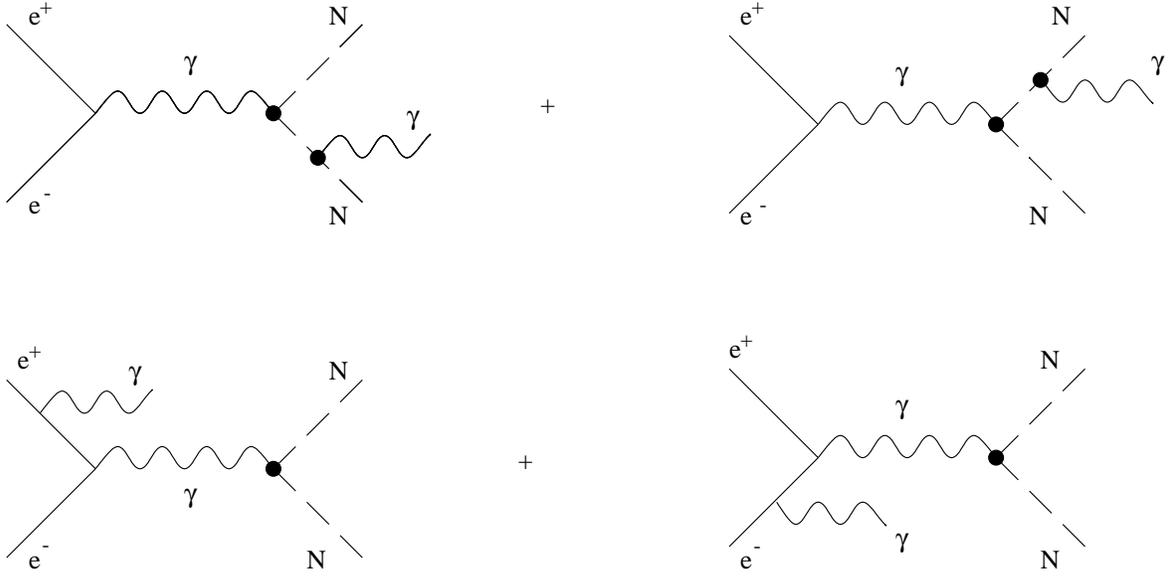}}  }
\caption{Feynman diagrams which contribute to the process $e^+e^- \rightarrow N \bar{N} \gamma$. }
\protect \label{figfeynman}
\end{figure}

The interaction of heavy neutrinos with the photon is $-ie \bar{N} D \sigma_{\mu \nu} \gamma_5
q^{\nu} N A^{\mu}$, where q is the momentum transfer. The EDM is defined to be $eD(q^2=0)$.  In this case, the
photon momenta at the relevant vertices are different, and are certainly not zero.   However, we presume that
$D(q^2)$ does not vary too rapidly with
$q^2$, as is the case in which the momentum transfer is not substantially greater than the masses of the
particles in the loop. The feynman diagrams are listed in Fig.
\ref{figfeynman}. The contribution due to the exchange of the Z boson is neglected here since it is numerically
small\cite{ftnt}. 

\begin{figure}
\centerline{ \epsfysize 4in \rotatebox{360}{\epsfbox{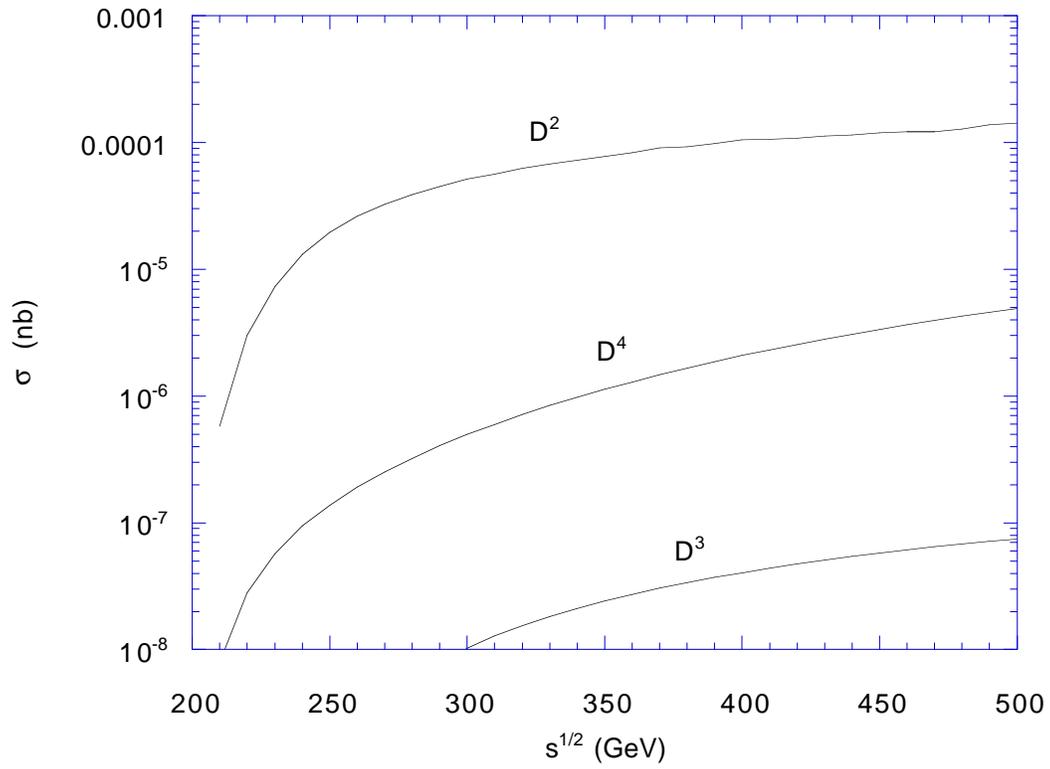}}  }
\caption{Different contributions to the total cross section for the process 
$e^+e^- \rightarrow N \bar{N} \gamma$
with the heavy neutrino mass of 100 GeV and $D=10^{-16} \ e$ cm.   For a positive EDM, the $D^3$ term is negative; the absolute
value is shown. }
\protect \label{figD2versusD4}
\end{figure}

The calculation is straightforward. It is clear
that the rate is the sum of terms proportional to $D^2$, $D^3$ and $D^4$.
The contributions proportional to $D^2$ , $D^3$ and  $D^4$ are presented separately in Fig.
\ref{figD2versusD4} for $D=10^{-16}$ e-cm.   For different $D$'s, these curves scale
appropriately.  One can see that for $D\ <\ 6\times 10^{-16}$ e-cm, expected
in realistic models, the
$D^2$ term dominates.  Because
of this scaling, instead of falling as
$1/s$, the cross section reaches a constant (it would grow as $s$ if the $D^4$ term dominated.  As
discussed above, however, it does not reach the level at which
unitarity is a big concern.  The total cross section is then given, to a good approximation, by the
$D^2$ term in Fig.
\ref{figD2versusD4}.

 The cross
section will be singular as the $|\cos \theta_{\gamma}|$ tends to 1. We have used cuts in the
photon energy and in the angle: $\frac{E_{\gamma}}{\sqrt{s}} \ge 0.1$ and $|\cos \theta_{\gamma}|
\le 0.94$.  For example, at $\sqrt{s}=500$ GeV, the cross section is about 
$1.5 \times 10^{-4}$ nb (and scales as $D^2$).  This is a very large event rate, corresponding to
about an event every few minutes (for $D=10^{-16}$ e-cm) at a linear collider with a luminosity of
$3
\times 10^{34} {\rm cm}^{-2}{\rm sec}^{-1}$.

\begin{figure}
\centerline{ \epsfysize 4in \rotatebox{360}{\epsfbox{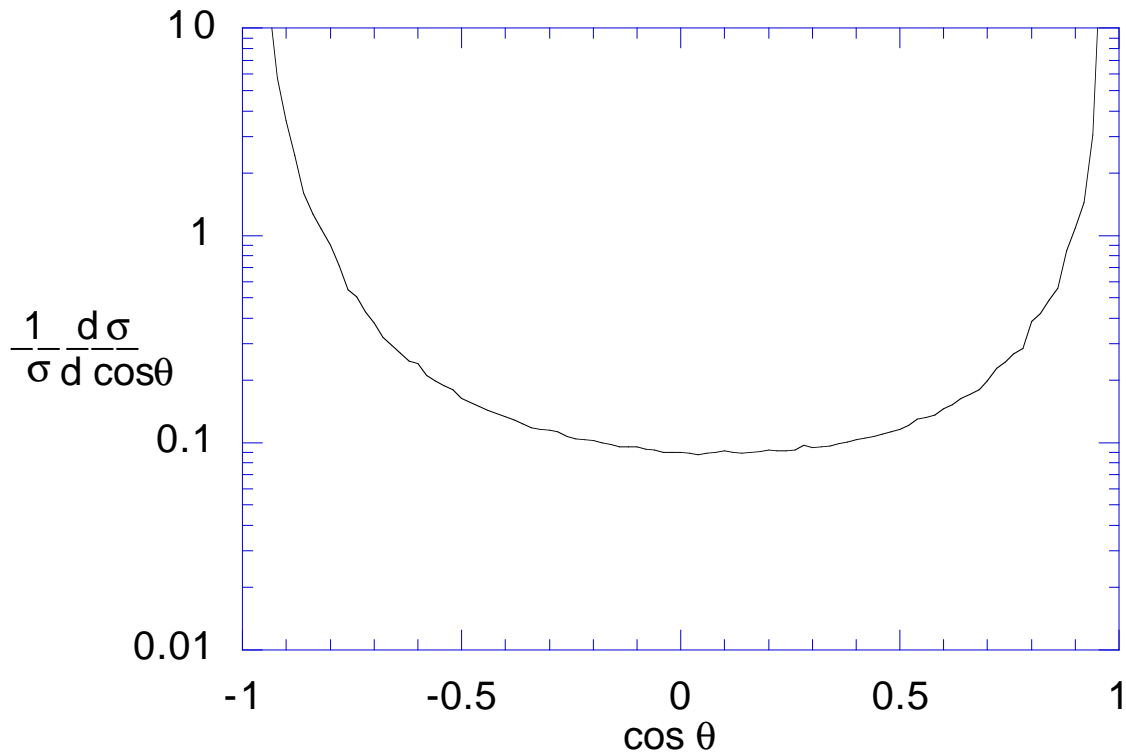}}  }
\caption{Angular distribution of the cross section  
with the heavy neutrino mass of 100 GeV and $D=10^{-16} \ e$ cm, and $s^{1/2}=500$ GeV.
  }
\protect \label{angularplot}
\end{figure}

\begin{figure}
\centerline{ \epsfysize 4in \rotatebox{360}{\epsfbox{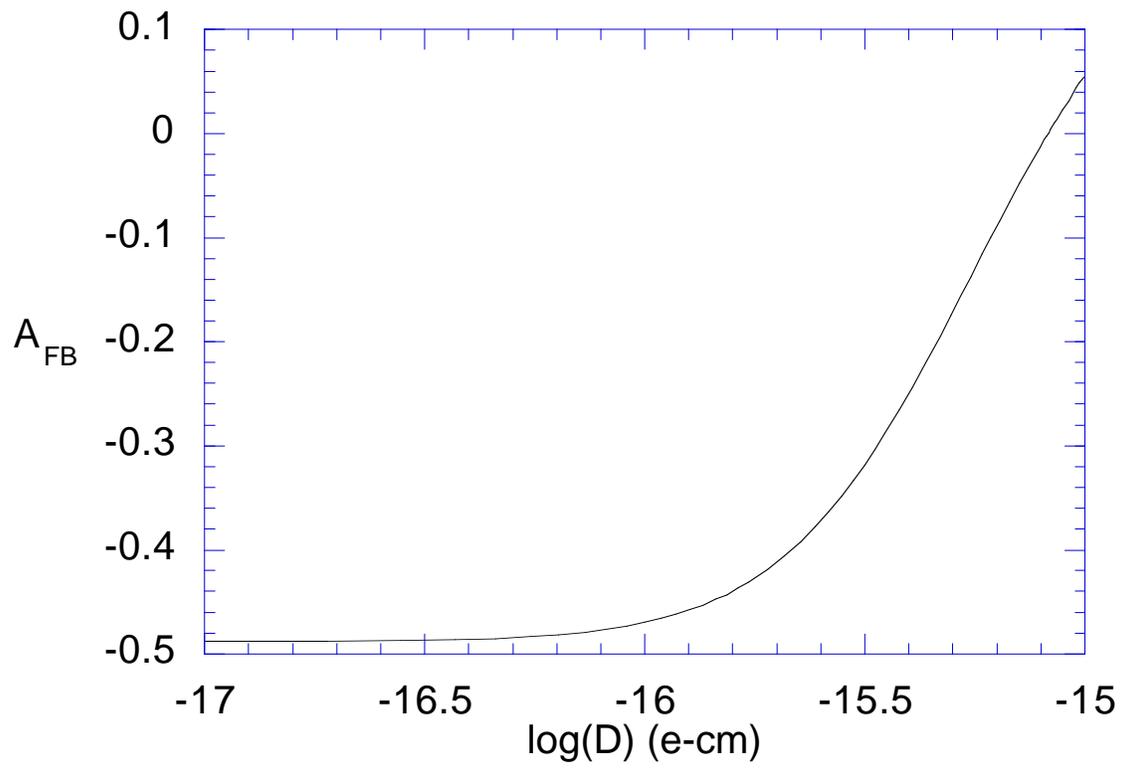}}  }
\caption{The forward-backward asymmetry as a function of the EDM.  }
\protect \label{angular2}
\end{figure}

The angular distribution is also important. It is shown in Fig. \ref{angularplot} for $D=10^{-16}$
e-cm, and the resulting forward-backward asymmetry is shown as a function of $D$ in Fig.
\ref{angular2}.  The angular distribution of the Standard Model is nearly forward-backward
symmetric,  but in this case there is a sizable asymmetry.   For much of the parameter space, the
asymmetry is $-0.48$.

Clearly, the first signature of an EDM for a heavy neutrino will be a very large event rate for
single photon events, with the forward-backward asymmetry measurement giving a confirmation of
this particular interpretation of such a large rate.

\section{Astrophysical and Cosmological Significance}

One of the greatest mysteries in cosmic ray astronomy concerns the existence of cosmic rays with energies above
the GZK bound\cite{gzk}.   A substantial number of cosmic rays with energies well above $10^{20}$ eV have been
observed, interacting in the upper atmosphere.  Yet a charged particle's mean free path through cosmic microwave
background, at these energies, is only about $50$ megaparsecs, and no sources of such high energy particles are
known to exist within this distance.   Neutral particles will not have this problem.  However, the fact that
the cosmic rays shower high up in the atmosphere implies that the cross-section must be of strong interaction
strength.  The only strongly interacting long-lived neutral particle is the neutron, yet its lifetime is too
short.  Neutrinos will not interact high in the atmosphere, if they have conventional weak interactions.

Several papers\cite{ralston} have considered the possibility that neutrino interactions might become strong at
very high energies, due to exchange of Kaluza-Klein gravitons\cite{ralston,graviton}.  One might wonder whether
heavy neutrinos with
a sufficiently large EDM might interact strongly enough in the upper atmosphere, but not interact strongly with
the microwave
background. Are these candidates for the ultra-high energy cosmic rays?    One can redo the calcuation described
above
concerning ionization loss in a detector, but now looking at the ultrarelativistic limit.  This is a tedious,
but straightforward calculation.  

In this case, the cross section can not be high enough.  Recall that the EDM is defined at $q^2=0$.  However,
here one would expect $q^2$ to be enormous.  Certainly, in the first few interactions, it will be much larger
than the mass-squared of the particles in the loop that produces a large EDM in the first place.  As a result,
the effective EDM will be much smaller, and the interaction will not be strong.  These heavy neutrinos will not
interact high in the atmosphere, and may not interact much more than conventional high energy neutrinos.

Are there any other potential astrophysical or cosmological effects?  One possibility concerns the generation
of the baryon asymmetry.  Suppose the heavy neutrino decays radiatively into light neutrinos.  Then the huge
EDM, which violates CP, may cause the decay rate into neutrinos to be different than the decay rate into
antineutrinos.  This would result in a large lepton asymmetry.  Since non-perturbative sphaleron interactions
can convert a lepton asymmetry into a baryon asymmetry, this would result in a baryon asymmetry.  The reader is
referred to the article of Riotto and Trodden\cite{riotto} for a nice review.  This possibility is currently
under investigation.

\section{Conclusions}

In this Article, we have explored the possibility that a heavy neutrino could have a huge electric dipole moment.  It has
been shown that plausible models exist that would lead to a dipole moment as large as $10^{-16}$ e-cm, and that this would
dramatically alter the electromagnetic properties of these neutrinos.  The first signature of such a neutrino would come from a
large enhancement of the single photon plus missing energy event rate.  Then, one could actually look for the ionization track of
these neutrinos in specially designed detectors.  Cross sections and angular distributions are calculated as a function of the
electric dipole moment.  While these neutrinos can not explain the ultra high energy cosmic rays, they may have interesting
implications for baryogenesis.

{\bf Acknowledgments}

We thank Mariana Frank as one of us (Nie) is now doing postdoctoral research 
under her supervision.   We also are extremely grateful to John Ralston, for repeatedly warning us that the EDM
at high energies would be softened, making it unlikely that these neutrinos could be the UHE cosmic rays. We
finally believed him.  We also thank Jack Kossler and Chris Carone for many useful discussions. 
This work is supported by the National Science Foundation grant number PHY-9900657.

\newpage

\def\oldprd#1#2#3{{\rm Phys. ~Rev. ~}{\bf D#1}, #3 (19#2)}
\def\newprd#1#2#3{{\rm Phys. ~Rev. ~}{\bf D#1}, #3 (20#2)}
\def\plb#1#2#3{{\rm Phys. ~Lett. ~}{\bf B#1}, #3 (19#2) }
\def\npb#1#2#3{{\rm Nucl. ~Phys. ~}{\bf B#1}, #3 (19#2) }
\def\prl#1#2#3{{\rm Phys. ~Rev. ~Lett. ~}{\bf #1}, #3 (19#2) }
\def\prl20#1#2#3{{\rm Phys. ~Rev. ~Lett. ~}{\bf #1}, #3 (20#2) }
\def\rep19#1#2#3{{\rm Phys. ~Rep. ~}{\bf #1}, #3 (19#2) }
\def\rep20#1#2#3{{\rm Phys. ~Rep. ~}{\bf #1}, #3 (20#2) }

\bibliographystyle{unsrt}

\begin{thebibliography}{99}
\bibitem{ftnt}
Note that the cross-section for mu-pair production via $\gamma$-exchange is bigger than that via $Z$-exchange, at high energies,
by a factor of $256\sin^4\theta_W\cos^4\theta_W$, which is approximately ten.
\end{thebibliography}

\end{document}